\documentclass[prl,aps,superscriptaddress,twocolumn,notitlepage,showpacs]{revtex4-1}
\usepackage{latexsym}
\usepackage{amsmath, dsfont}
\usepackage{amssymb}
\usepackage{graphicx}
\usepackage{caption}
\usepackage{physics}
\usepackage{subfigure}
\usepackage{float}
\usepackage{calrsfs}
\usepackage{mathrsfs}
\usepackage{notes2bib}
\usepackage{color}
\usepackage{txfonts}
\usepackage[justification=centering,
            format=plain]{caption}

\renewcommand{\raggedright}{\leftskip=0pt \rightskip=0pt plus 0cm}

\begin{document}

\title{Enhanced sensing of weak anharmonicities through coherences in dissipatively coupled anti-PT symmetric systems}

\author{Jayakrishnan M. P. Nair}
\email{jayakrishnan00213@tamu.edu}
\affiliation{Institute for Quantum Science and Engineering, Department of Physics and Astronomy,Texas A$\&$M University, College Station, TX 77843, USA}

\author{Debsuvra Mukhopadhyay}
\email{debsosu16@tamu.edu}
\affiliation{Institute for Quantum Science and Engineering, Department of Physics and Astronomy,Texas A$\&$M University, College Station, TX 77843, USA}

\author{G. S. Agarwal}
\email{girish.agarwal@tamu.edu}
\affiliation{Institute for Quantum Science and Engineering, Department of Physics and Astronomy,Texas A$\&$M University, College Station, TX 77843, USA}
\affiliation{Department of Biological and Agricultural Engineering, Texas A$\&$M University, College Station, TX 77843, USA}

\date{\today}

\begin{abstract}
%We analyze how an interplay between dissipative coupling and the phenomenon of vacuum induced coherence (VIC) in a two-mode framework permits a fine-grained estimation of weak Kerr nonlinearity. We establish this by considering an optomagnonic apparatus that enables a highly sensitive detection of weakly anharmonic oscillations of magnons in a Yttrium Iron Garnet (YIG) sphere. The sphere is coupled dissipatively to a single-mode cavity through an interposing fiber waveguide. By tamping down environmental losses to achieve VIC and driving the YIG by a microwave laser close to resonance, we substantiate the feasibility of inducing considerably high magnetization which exhibits an anomalous surge with diminishing nonlinearity. The purely dissipative nature of our model contrasts with many of the conventional systems which utilize an antisymmetric gain-loss profile for sensing applications.
In the last few years, the great utility of PT-symmetric systems in sensing small perturbations has been recognized. Here, we propose an alternate method relevant to dissipative systems, especially those coupled to the vacuum of the electromagnetic fields. In such systems, which typically show anti-PT symmetry and do not require the incorporation of gain, vacuum induces coherence between two modes. Owing to this coherence, the linear response acquires a pole on the real axis. We demonstrate how this coherence can be exploited for the enhanced sensing of very weak anhamonicities at low pumping rates. Higher drive powers ($\sim 0.1$ W), on the other hand, generate new domains of coherences. Our results are applicable to a wide class of systems, and we specifically illustrate the remarkable sensing capabilities in the context of a weakly anharmonic Yttrium Iron Garnet (YIG) sphere interacting with a cavity via a tapered fiber waveguide. A small change in the anharmonicity leads to a substantial change in the induced spin current.
\end{abstract}

\maketitle
%\pagenumbering{roman}

%\newpage

%\pagenumbering{arabic}

%\section{Introduction}
%The manipulation of coherent coupling in hybrid quantum systems has been at the heart of quantum optics and information processing for several years. In particular, strong light-matter interaction constitutes the building block for a myriad of sophisticated quantum devices such as sensors \cite{gil2010nanomechanical, zhu2010chip, he2011detecting, vollmer2012review, PhysRevLett.108.120801}.
 In the modern world with proliferating technological advances, sensing is of fundamental importance, with far-reaching applications \cite{gil2010nanomechanical, zhu2010chip, he2011detecting, vollmer2012review, PhysRevLett.108.120801} across various scientific disciplines, with adoptions as particle sensors, motion sensors and more. Both semiclassical and quantum phenomena provide us with a wide range of techniques to attain remarkable efficacy in sensing operations. Over the past decade, parity-time (PT) symmetric \cite{PhysRevLett.80.5243, bender2002generalized, PhysRevLett.89.270401, El-Ganainy:07} systems with a balanced loss and gain have been revealed to possess an enormous potential in boosting sensitivity. Thus, the non-Hermitian degeneracies known as exceptional points (EPs) in PT-symmetric systems \cite{chang2014parity, PhysRevA.99.053806, heiss2012physics, xu2016topological, RevModPhys.87.61}  have rendered a new avenue to engineer augmented response in an open quantum system \cite{chen2017exceptional, hodaei2017enhanced, PhysRevLett.123.213901, lin2016enhanced, PhysRevLett.112.203901, PhysRevApplied.5.064018, miri2019exceptional, PhysRevLett.123.237202, zeng2019enhanced, PhysRevA.93.033809}. Some recent experiments include the demonstration of enhanced sensitivity in microcavities near EPs \cite{chen2017exceptional} and the observation of higher-order EPs in a coupled-cavity arrangement \cite{hodaei2017enhanced}. While this is a truly remarkable development based on a balanced gain-loss distribution, one would like to examine the possibilities of newer sensing methodologies, where one can avoid the use of gain and yet, attain extremely high sensitivity. All of this would be without using quantum resources such as entangled photons, the innate potential of which would further boost the sensitivity.

In this letter, we demonstrate a new physical basis for enhanced sensing without utilizing a commensurate gain-loss profile. We consider dissipatively coupled systems where the coupling is produced via interaction with the vacuum of the electromagnetic field \cite{agarwal1974quantum}. Such systems have the novel property that vacuum induces coherence between two modes. The phenomenon of vacuum induced coherence (VIC) has been the subject of intense activity \cite{agarwal2000anisotropic, jha2015metasurface, lassalle2020long, kornovan2019noninverse, scully2011quantum, kiffner2010vacuum, keitel1999narrowing, zhou1997quantum, paspalakis1998phase, paspalakis1998fluorescence, heeg2013vacuum, dorfman2013photosynthetic, PhysRevLett.104.207701, PhysRevA.84.053818, dodin2019light} with applications ranging from heat engines \cite{scully2011quantum}, nuclear gamma ray transmission \cite{heeg2013vacuum} to photosynthesis \cite{dorfman2013photosynthetic} and molecular isomerization in vision \cite{dodin2019light}. In a system with strong VIC, one of the eigenvalues characterizing its dynamics moves to the real axis. We demonstrate the great utility of this key property to the sensing of extremely weak nonlinearities which are, otherwise, difficult to detect. This new paradigm is applicable generally to a wide class of systems encountered across various scientific disciplines. Examples include quantum dots coupled to plasmonic excitations in a nanowire \cite{wei2011quantum}, superconducting transmon qubits \cite{PhysRevA.76.042319}, quantum emitters coupled to meta materials \cite{agarwal2000anisotropic, jha2015metasurface, lassalle2020long, PhysRevB.95.075412}, optomechanical systems \cite{PhysRevA.98.023841}, hybrid magnon-photon systems \cite{PhysRevLett.121.137203} and more.

We show explicit results on enhanced sensing by employing the VIC paradigm in an anti-PT symmetric configuration to the detection of very weak magnetic nonlinearities in a YIG sphere coupled to a cavity \cite{PhysRevLett.121.137203, PhysRevB.99.134426,  PhysRevLett.123.127202, PhysRevB.101.064404, yao2019coherent, wang2020dissipative, PhysRevX.5.021025, PhysRevLett.123.227201, PhysRevLett.113.156401, PhysRevLett.113.083603, tabuchi2015coherent, PhysRevLett.125.117701, PhysRevB.94.224410}. This system is specifically chosen in view of the ongoing experimental activities. YIGs are endowed with high spin density and the collective motion of these spins are embodied in the form of quasiparticles named magnons. Dissipative coupling using YIGs has been observed in a multitude of settings, involving, for instance, a Fabry-perot cavity \cite{PhysRevLett.121.137203} or a coplanar waveguide \cite{PhysRevB.99.134426}. Note that under most circumstances, weak nonlinearities of the order of nHZ would require immense drive power to be detected in experiments. However, a dissipatively coupled system affords a prodigious response in the magnetization of the YIG which goes up spectacularly with the weakening strength of nonlinearity. That our system does not require the administration of gain to draw out such a divergent response underpins its utility in sensing applications. Upon ramping up the drive power, anharmonic effects are strongly reinforced and new domains of VIC are brought to light, with peaks in the response function corresponding to strong linewidth suppression.

We start off by considering the general model for a two-mode anharmonic system, which is pertinent to a wide range of physical systems. This is characterized by a Hamiltonian
\begin{equation}
\begin{split}
{H}/\hbar= \omega_{a}a^{\dagger}a+\omega_{b}b^{\dagger}b+g(ab^{\dagger}+a^{\dagger}b)&\\+U(b^{\dagger 2}b^2)+i\Omega (b^{\dagger}e^{-i \omega_d t}-b e^{i \omega_dt}),
\end{split}
\end{equation}
where $\omega_{a}$ and $\omega_{b}$ denote the respective resonance frequencies of the uncoupled modes $a$ and $b$, and $g$ constitutes the coherent hermitian coupling between them. The parameter $U$ is a measure of the strength of anharmonicity intrinsic to the mode $b$, which is driven externally by a laser at frequency $\omega_d$. The quantity $\Omega$ represents the Rabi frequency. In addition, these modes could be interfacing with a dissipative environment. Dissipative environments in an open quantum system fall roughly under two classifications - one, where the modes are coupled independently to their local heat baths, and another, where a common reservoir interacts with both, as depicted in figure (1).

\begin{figure}
 \captionsetup{justification=raggedright,singlelinecheck=false}
 \centering
   \includegraphics[scale=0.48]{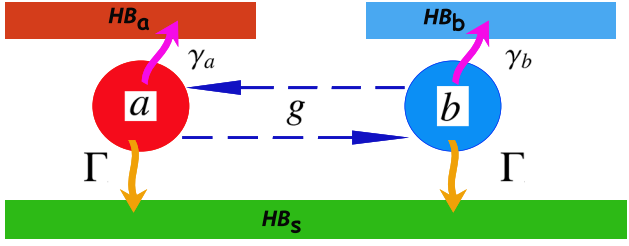}
\caption{Schematic of a general two-mode system dissipatively coupled through a waveguide. $\gamma_{a(b)}$ and $\Gamma$ describe decay into the surrounding (local heat bath) and coupling to the fiber (shared bath) respectively. Optimal effect of VIC can be realized in the limit of $\gamma_{a(b)}\ll \Gamma$.}
\label{sch}
\end{figure}

A complete description of the two-mode system, in terms of its density matrix $\rho$, is provided by the master equation \cite{agarwal1974quantum} 
\begin{equation}
\frac{\dd \rho}{\dd t}=-\frac{i}{\hbar}[{H},\rho]+\gamma_{a}\mathcal{L}(a)\rho+\gamma_{b}\mathcal{L}(b)\rho+2\Gamma\mathcal{L}(c)\rho,
\end{equation}
where $\gamma_{a}$ and $\gamma_{b}$ are, respectively, the intrinsic damping rates of the modes, induced by coupling with their independent heat baths. The parameter $\Gamma$ introduces coherences, and the Liouvillian operator $\mathcal{L}$ is defined by its action $\mathcal{L}(\sigma)\rho=2\sigma\rho \sigma^{\dagger}-\sigma^{\dagger}\sigma\rho-\rho \sigma^{\dagger}\sigma$. Assuming symmetrical couplings of the modes to the common reservoir we have the relation $c=  \frac{1}{\sqrt{2}}(a+ b)$. The mean value equations for $a$ and $b$ are obtained to be
 \begin{equation}
 \begin{split}
 \begin{pmatrix}
\dot{a} \\
\dot{b}
\end{pmatrix} =-i\mathscr{H}\begin{pmatrix}
a \\
b
\end{pmatrix}-2iU(b^{\dagger}b)\mathscr{R}\begin{pmatrix}
a \\
b
\end{pmatrix}+\Omega e^{-i\omega_dt}\begin{pmatrix}
0 \\
1
\end{pmatrix},
\end{split}
 \end{equation}
where $\mathscr{H}=\begin{pmatrix}
\omega_{a}-i(\gamma_{a}+\Gamma) & g-i\Gamma \\
g-i\Gamma & \omega_{b}-i(\gamma_{b}+\Gamma) 
\end{pmatrix}$, $\mathscr{R}=\begin{pmatrix}
0 & 0 \\
0 & 1 
\end{pmatrix}$, and the notation $\expval{.}$ has been dropped for conciseness. In dealing with the nonlinear term, we have taken recourse to the mean-field approximation $\expval{\mathcal{X}_1\mathcal{X}_2}=\expval{\mathcal{X}_1}\expval{\mathcal{X}_2}$ for any two operators $\mathcal{X}_1$ and $\mathcal{X}_2$. A canonical transformation of the form $(a,  b)$ $\rightarrow$ $e^{-i\omega_dt}(a, b)$ stamps out the time dependence on the final term in (3) and translates $\mathscr{H}$ to $\mathscr{H}-\omega_d\mathds{1}$ without tampering with the nonlinear term, where $\mathds{1}$ is the $2\cross 2$ identity matrix. This transformation takes us to the frame of the applied laser frequency.

Before proceeding with a generalized treatment, let us first deconstruct the linear dynamics, i.e. when $U$ is dropped. Defining the detuning parameters $\Delta_{a}=\omega_{a}-\omega_d$, $\Delta_{b}=\omega_{b}-\omega_d$, we have the eigenvalues of $\mathscr{H}$ given by $\lambda_{\pm}=\omega_d+{\Delta_0}-i(\gamma_0+\Gamma)\pm\sqrt{(\Delta-i\gamma_{ab})^2+(g-i\Gamma )^{2}}]$, with ${\Delta}_0=(\Delta_{a}+\Delta_{b})/2$, $\Delta=(\Delta_{a}-\Delta_{b})/2$, ${\gamma_0}=(\gamma_{a}+\gamma_{b})/2$ and $\gamma_{ab}=(\gamma_{a}-\gamma_{b})/2$. Contingent on the stability condition $\Im(\lambda_{\pm})<0$, which averts exponential amplification, the steady-state solutions for the mean values $\mathscr{O}^{(0)}=\expval{O}$  $(\mathscr{O}^{(0)}=\mathscr{A}, \mathscr{B}; O=a, b)$ would unfold as
 \begin{align}
\mathscr{A}&=\frac{-g+i\Gamma }{(\omega_d-\lambda_{+})(\omega_d-\lambda_{-})}\Omega\notag\\
\mathscr{B}&=\frac{\Delta_{b}-i(\gamma_b+\Gamma)}{(\omega_d-\lambda_{+})(\omega_d-\lambda_{-})}\Omega
\end{align}
It makes for a straightforward inference that the resonant responses of these steady-state amplitudes pertain, in principle, to the eigenfrequencies of $\mathscr{H}$. In the most generic setting, since the eigenmodes have finite linewidths, it is not possible to chart a real parameter trajectory through these roots. However, there exist leeways for certain classes of systems, which we make clear in the following discussion.

\begin{figure}
 \captionsetup{justification=raggedright,singlelinecheck=false}
 \centering
   \includegraphics[scale=0.48]{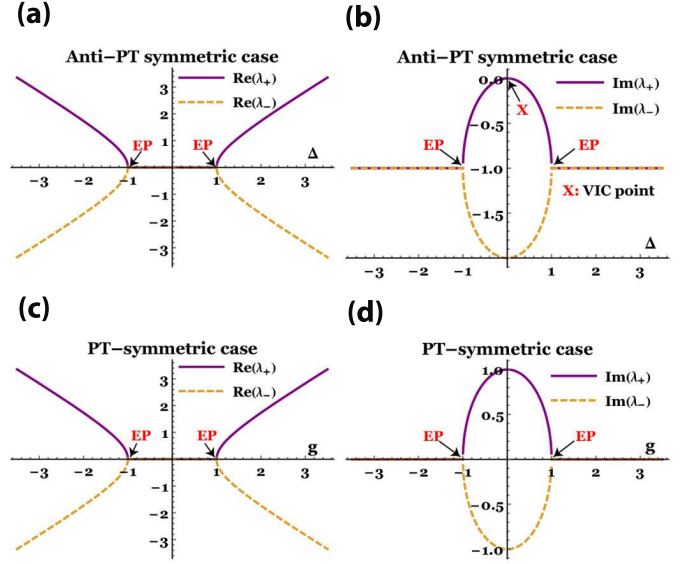}
\caption{a), b) Eigenfrequencies and linewidths for an anti-PT symmetric system, plotted against a variable $\Delta$, viewed in units of $\gamma_{0}=\Gamma$. While EPs emerge at $\Delta=\pm \gamma_0$, the VIC-induced linewidth suppression corresponds to $\Delta=0$. c), d) Analogous plots for the PT symmetric system, against the coupling strength $g$, in units of $\gamma_{ab}$, at $\Delta=0$. EPs are found at $g=\pm\gamma_{ab}$. In all of these cases, the eigenvalues refer to the Hamiltonian $\mathscr{H}-\omega_d\mathds{1}$.}
\label{sch}
\end{figure}

The generic $2\cross 2$ matrix $\mathscr{H}$ encompasses two distinct subtypes: (i) coherently coupled systems with $\Gamma=0$, and (ii) dissipatively coupled systems with $g=0$. Two special symmetries can be realized within these folds, namely {PT}-symmetry (allowed by (i)), which obey $(\hat{PT})\mathscr{H}(\hat{PT})=\mathscr{H}$, and anti-{PT} symmetry (allowed by (ii)), with $(\hat{PT})\mathscr{H}(\hat{PT})=-\mathscr{H}$. An anti-{PT} symmetric realization of mode hybridization can be pinned down by the parameter choices $\Delta_{0}=\gamma_{ab}=g=0$ and $\Gamma\neq 0$. Clearly, while the modes are oppositely detuned in character, both of them retain their lossy nature. Here, it makes sense to switch to the rotating frame of the laser, where $\mathscr{H}$ is substituted by $\mathscr{H}-\omega_d\mathds{1}$. The shifted eigenvalues would be obtained as $-i(\gamma_{0}+\Gamma)\pm\sqrt{\Delta^2-\Gamma^2}$ for $\abs{\Delta}>\Gamma$ and $-i(\gamma_{0}+\Gamma)\pm i\sqrt{\Gamma^2-\Delta^2}$ (broken anti-PT) for $\abs{\Delta}<\Gamma$.  The behavior of the real and imaginary parts of these eigenvalues is provided in figure 2 (a), (b). As long as the stability criterion is fulfilled, the responses in (4) are inversely related to $(\omega_d-\lambda_{+})(\omega_d-\lambda_{-})=-\gamma_0(2\Gamma+\gamma_0)-\Delta^{2}$. The broken anti-PT phase brings in real singularities at $\omega_d=\frac{1}{2}(\omega_a+\omega_{b})$ in the limit $\gamma_{0} \rightarrow 0$, which is evidenced by the resonant inhibition in the imaginary part of $\lambda_+$, as marked by the point X in figure 2 (b). The extreme condition $\gamma_{0}=0$ holds when none of the modes suffers spontaneous losses to its independent surrounding, all the while interacting with the mediating reservoir. Therefore, by harnessing dissipative coupling between two modes and optimizing the effect of VIC, we observe a precipitous divergence in the linear response under steady-state conditions. 

%The {PT}-symmetric configuration for coherently coupled systems is conformable with the parameter structure $\Delta=\gamma_0=\Gamma=0$. Therefore, injection of gain constitutes a necessary precursor for this symmetry. With the shifted eigenvalues given as ${\Delta}_0\pm i\sqrt{\gamma_{ab}^2-g^2}$ for $\abs{\gamma_{ab}}>g$ and as ${\Delta}_0\pm \sqrt{g^2-\gamma_{ab}^2}$ for $\abs{\gamma_{ab}}<g$, it follows that $\abs{\gamma_{ab}}=g$ represents a transition point signifying the passage from PT-symmetric hybridization to broken-PT phase (figure 2. c), d)). More importantly, this bifurcation point, which defines the EP, introduces real singularities and strongly accentuated resonances in the mode amplitudes at $\omega_a=\omega_{b}=\omega_{d}$.  

The {PT}-symmetric configuration for coherently coupled systems is conformable with the parameter structure $\Delta=\gamma_0=\Gamma=0$. The constraint $\gamma_b=-\gamma_a$ implies that a loss in mode $a$ must be offset by a commensurate gain in mode $b$. With the corresponding eigenvalues given as ${\Delta}_0\pm i\sqrt{\gamma_{ab}^2-g^2}$ for $\abs{\gamma_{ab}}>g$ and as ${\Delta}_0\pm \sqrt{g^2-\gamma_{ab}^2}$ for $\abs{\gamma_{ab}}<g$, it follows that the transition point $\abs{\gamma_{ab}}=g$, which defines the EP, introduces real singularities and strongly accentuated resonances in the mode amplitudes at $\omega_a=\omega_{b}=\omega_{d}$. Figure 2 (c), (d) depicts the eigenvalue structure in the PT-symmetric case.

Next, we illustrate the importance of the condition $\Im(\lambda_{+})\rightarrow 0$ in the context of the nonlinear response observed in the system. The nonlinear behavior depends on the intrinsic symmetry properties of the matrix $\mathscr{H}$. Specifically, the extraordinary response achievable in anti-PT symmetric models yields a convenient protocol for the fine-grained estimation of weak anharmonicity. We restrict our focus to the anti-PT symmetric framework because it circumvents the possible difficulties of gain fabrication. We now consider a full treatment of Eq. (3) by factoring in the effect of $U$. In the rotating frame, upon choosing $\Delta_{a}=-\Delta_{b}=\delta/2$, and $\gamma_{a}=\gamma_{b}={\gamma_{0}}$, Eq. (3) leads to the modified steady-state relations:
\begin{align}
-(i\delta/2 + {\gamma_0}+\Gamma)a-\Gamma b&=0\nonumber\\[0.15cm]
-(-i\delta/2 + {\gamma_0}+\Gamma)b-2iU|b|^{2}b-\Gamma b+\Omega&=0
\label{QLE}
\end{align} 
Defining $\gamma={\gamma_0}+\Gamma$ and eliminating $a$, the intensity $x=|b|^{2}$ is found to satisfy a cubic relation
\begin{align}
& \frac{\beta^{2}}{\gamma^{2}+(\delta/2)^{2}}x-\frac{2U\beta \delta}{\gamma^{2}+(\delta/2)^{2}}x^{2}+4U^{2}x^{3}=I,
\label{spinc}
\end{align}
where $\beta=\Gamma^{2}-\gamma^{2}-(\delta/2)^{2}$ and $I=\Omega^{2}$. Eq. (6) can entail a bistable response under the condition $U\delta<0$ and $\delta^{2}>12\gamma^{2}$. However, throughout this manuscript, we operate at adequately low drive powers to ward off bistable signature. Now, in the limit ${\gamma_0}\rightarrow 0$ and $\delta\rightarrow 0$, $\beta$ becomes vanishingly small, and the first two terms in Eq.  (6) recede in importance, for a given Rabi frequency $\Omega$. Consequently, in the neighborhood of $\delta=0$, the response becomes highly sensitive to variations in $U$. To be more precise, for sufficiently low values of the detuning, the response mimics the functional dependence $x\approx (I/4U^{2})^{1/3}$. A tenfold decrease in $U$, therefore, scales up the peak intensity of $b$ by a factor of $4.64$. In this context, it is useful to strike a correspondence with the sensitivity in eigenmode splitting around an EP which is typically employed in PT symmetric sensing protocols \cite{chen2017exceptional, hodaei2017enhanced, PhysRevLett.112.203901}. For two mode systems, where the EP is characterized by a square root singularity, this splitting $\delta \omega$ scales as the square root of the perturbation parameter $\epsilon$ implying a sensitivity that goes as $\abs{\frac{\delta \omega}{\delta \epsilon}}\propto \abs{\epsilon}^{-1/2}$. However, in our setup, the sensitivity to $U$ in the response is encoded as $\abs{\frac{\delta x}{\delta U}}\propto \abs{U}^{-5/3}$.

The importance of the above result in the context of sensing is hereby legitimized for dissipatively coupled systems. Guided by the recent experiments on dissipatively coupled hybrid magnon-photon systems \cite{PhysRevLett.121.137203, PhysRevB.99.134426, PhysRevLett.123.127202, PhysRevB.101.064404, yao2019coherent, wang2020dissipative}, we apply these ideas to the specific example of Kerr nonlinearity in a YIG sample \cite{PhysRevB.94.224410}. However, the bulk of these works have restricted their investigations to the linear domain. Here, we transcend this restriction and study the nonlinear response to an external drive. We consider an integrated apparatus comprising an optical cavity and a YIG sphere, both interfacing with a one-dimensional waveguide. The direct coupling between the cavity and the magnon modes can be neglected. However, the interaction with the waveguide would engender an indirect coupling between them. In order to excite the weak Kerr nonlinearity of the YIG sphere, a microwave laser is used to drive the spatially uniform Kittel mode. The full Hamiltonian in presence of the external drive can be cast exactly in the form of Eq. (1), with $b$ superseded by the magnonic operator $m$ \cite{key},
\begin{equation}
\begin{split}
H_{\text{eff}}/\hbar = & \omega_a a^{\dagger}a + \Big[\omega_{m} m^{\dagger}m + U (m^{\dagger 2}m^2)\Big] +\\& i\Omega\left(m^{\dagger} e^{-i\omega_{\text{d}}t} - m e^{i\omega_{\text{d}}t}\right).
\end{split}
\end{equation}

\begin{figure}
 \captionsetup{justification=raggedright,singlelinecheck=false}
 \centering
   \includegraphics[scale=0.43]{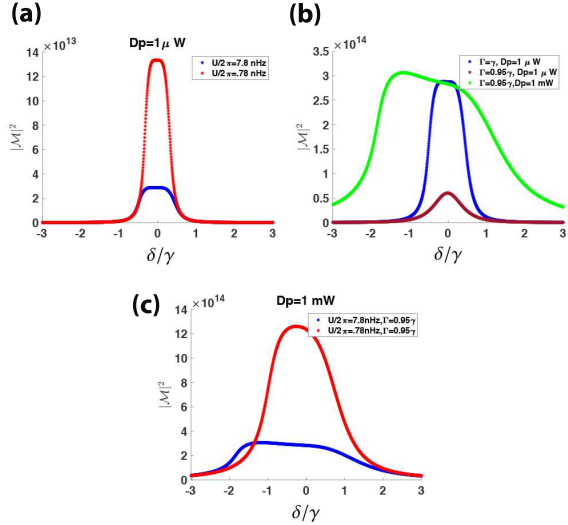}
\caption{a) The spin current plotted against $\delta$ at two different nonlinearities; b) spin currents away from the VIC condition, compared against the lossless scenario, at different drive powers and at $U/2\pi=7.8$ nHz - for ease of comparison, the blue and maroon curves have been scaled up by $10$; c) contrasting responses observed at a drive power of $1$ $mW$ for two different strengths of nonlinearity.}
\label{sch}
\end{figure}

As discussed earlier, the mediating effect of the waveguide is reflected as a dissipative coupling between the two modes, which instills VIC into the system. With the anti-PT symmetric choices $\Delta_{a}=-\Delta_{m}=\delta/2$, $\gamma_{a}=\gamma_{b}={\gamma_0}$, and the redefinition ${\gamma_0}+\Gamma=\gamma$, we recover Eq. (6) in the steady state, with the obvious substitution $b\rightarrow m$ and $x=|{b}|^{2}$ denoting the spin current response. We now expound the utility of engineering a lossless system in sensing weak Kerr nonlinearity. To that end, we zero in on the parameter subspace $\Gamma=\gamma=2\pi\cross 10$ MHz. Since $\beta=\delta^{2}/4$, the contributions from the first two terms in Eq. (6) taper off as resonance is approached. As outlined earlier, we find that for all practical purposes, the nonlinear response can be approximated as $x\approx (I/4U^{2})^{1/3}$ in the region $\delta/2\pi<1$ MHz, which demonstrates its stark sensitivity to $U$. A lower nonlinearity begets a higher response, as manifested in figure 3 (a), where plots of $x$ against $\delta$ are studied at differing strengths of the nonlinearity. Even at $D_p=1$ $\mu W$, we observe a significant enhancement in the induced spin current of the YIG around $\delta=0$. The result is a natural upshot of the VIC-induced divergent response in an anti-PT symmetric system in the linear regime. Quite conveniently, the inclusion of nonlinearity dispels the seemingly absurd problem of a real singularity noticed in the linear case. If $\Gamma < \gamma$, a strong quenching in the response is observed, as depicted in figure 3 (b). The sensitivity to variations in $U$ also incurs deleterious consequences. Nevertheless, we can counteract this decline by boosting the drive power. A drive power close to $1$ $mW$ can bring back the augmented response and the pronounced sensitivity to $U$ (figure 3 (c)). This mechanism can serve as an efficient tool to sense small anharmonicities present in a system. The fact that even a minute pump power in the range of $1$ $\mu W$ to $1$ $mW$ generates substantial enhancement in the spin current makes it all the more robust. 

As evident from the preceding discussion, it is imperative that the waveguide-mediated coupling overshadows the effect of spontaneous emissions. Moreover, the protocol hinges on the anti-PT symmetric character and eigenmodes of $\mathscr{H}$, which largely control the dynamics at low drive powers. At larger drive powers ($\sim0.1$ $W$), the nonlinear correction in (3) becomes important and activates new coherences. A theoretical explanation of this phenomenon can be spelled out by linearizing the dynamics of the mode $m$ about its steady-state value ensuing from (6). This linearization yields a higher-dimensional eigensystem, portrayed in figure 4 (a). The new coherences are closely correlated with the extreme linewidth narrowing manifested in the higher-dimensional model. Figure 4 (b) exemplifies new VIC-induced peaks that emerge even when spontaneous decays become comparable to the dissipative coupling. Details on this calculation are provided in the supplementary material \cite{key}. 

\begin{figure}
 \captionsetup{justification=raggedright,singlelinecheck=false}
 \centering
   \includegraphics[scale=0.48]{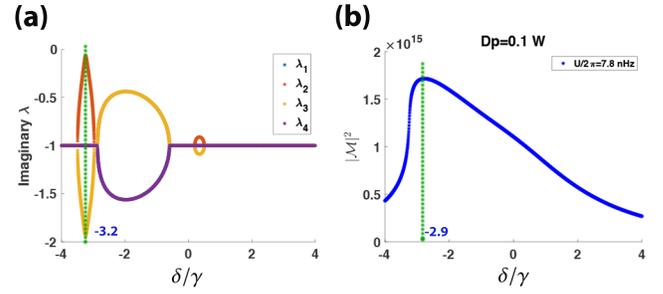}
\caption{Nonlinearity induced coherences and higher dimensionality: (a) Imaginary parts of the eigenmodes of an effectively four-dimensional system at a drive power of $0.1$ $W$ - around $\delta/\gamma=-3.2$, an extreme linewidth narrowing is observed; b) the spin current response, with a peak near to $\delta/\gamma=-2.9$, highly skewed due to a stronger drive.}
\label{sch}
\end{figure}
 
%\section{Level Attraction and Transmission:}

%\section{Conclusions:}
In summary, we have proposed an optical test bed that shows enhanced sensitivity to Kerr nonlinearity in the mode responses to an external field, hence qualifying it as a prototypical agency to gauge the strength of anharmonic perturbations under optimal conditions. The physical origin of this peculiar behavior lies in an effective coupling induced between the cavity and the magnon modes in the presence of a shared ancillary reservoir. Such a coupling is purely an artifact of third-party mediation, and is dissipative in nature as the two modes synergistically drive up the energy being channeled into the interposing reservoir. Optimal results vis-\`a-vis the estimation of nonlinearity are obtained when VIC strongly dominates, i.e when spontaneous emissions from the modes to the surrounding environments become negligible in comparison to the waveguide-mediated coupling. Since dissipatively coupled systems do not require the synthetic introduction of gain for efficient sensing, our setup offers a clear edge over PT-symmetric systems which rely on a balanced trade-off between gain and loss. At higher drive powers, we observe skewed VIC peaks as a testimony to strongly anharmonic responses, even when decays into the environment become significant. These nonlinearity-induced VICs could bear relevance in other contexts and merits further investigation. Although, to provide numerical estimates, our analysis has been tailored to demonstrate the sensitivity in the context of YIG oscillations, the essence of our assessment would be applicable to any two-mode nonlinear system. 

\section{Acknowledgments}
The authors gratefully acknowledge the support of The Air Force Office of Scientific Research [AFOSR award no FA9550-20-1-0366], The Robert A. Welch Foundation [grant no A-1243] and the Herman F. Heep and Minnie Belle Heep Texas A$\&$M University endowed fund. GSA thanks Dr. C. M. Hu for discussions on the dissipative coupling and for sharing his data with us.
\bibliography{main}
\end{document}